\documentclass[epj]{webofc}
\usepackage[varg]{txfonts}   
\usepackage{color}
\usepackage{slashed}

\definecolor{darkgreen}{rgb}{0,0.5,0}
\woctitle{MESON2018 - the 15$^\textrm{th}$ International Workshop on Meson Physics}
\begin{document}
\selectlanguage{english}
\title{News from the Lattice: example of the pion distribution amplitude}

\author{Piotr Korcyl\inst{1,2}\fnsep\thanks{\email{piotr.korcyl@uj.edu.pl}
}}

\institute{M. Smoluchowski Institute of Physics, Jagiellonian University, ul. \L ojasiewicza 11, 30-348 Krak\'ow, Poland
\and
Institut f\"ur Theoretische Physik, Universit\"at Regensburg, D-93040 Regensburg, Germany
}

\abstract{
In this contribution I present a brief summary of recent progress in selected Lattice QCD techniques and subsequently illustrate them using the example of the current calculations of moments of meson distribution amplitude as well as of a direct evaluation of meson distribution amplitude.
}
\maketitle
It is clearly an impossible task to summarize the progress in numerical Monte Carlo simulations of Quantum Chromodynamics 
of the last two years in less than half an hour. Many great advacements and breakthroughs, such as the study of resonances \cite{resonances}, three-body scattering \cite{scattering}, 
the study of nuclear forces \cite{nuclear}, and many more cannot be given proper credit in such a short time. What I will try to do instead, is that 
I will briefly mention four recent algorithmic advancements and illustrate their impact on the study of just a single observable, 
namely the pion distribution amplitude.

Pion distribution amplitude is one of the many hadron structure functions which describe the content of hadrons, in that particular case of the pion. Structure 
functions are still under debate, even 50 years after the postulation of Quantum Chromodynamics. On the experimental side they were and are
extensively investigated by experiments on past and current accelerators such as HERA or LHC. They are also the main focus of next generation facilities 
such as the Electron-Ion Collider planned to be constructed in the US. On the theory side little doubt remains that strong interactions are described by 
Quantum Chromodynamics, however estimating predictions for low-energy observables is still problematic due to the strongly coupled nature of that theory.
One of the most reliable tools to study QCD nonperturbatively are numerical Monte Carlo simulations, lattice QCD, which I briefly describe in the next section.
\section{Lattice Quantum Chromodynamics: theoretical 'experiments'}
Lattice QCD is one of the few formulations of Quantum Chromodynamics which is fully nonperturbative and allows to estimate QCD predictions with 
all the systematic uncertainties under control. It is a numerical framework in which one can perform theoretical 'experiments' with his favorite 
quantum field theory. Various particles or excitations can be placed and handled within the simulated, Euclidean volume of space-time with the help of quark 
field operators or appropriate, more complicated, interpolators. All particle properties can be extracted from euclidean correlation functions. 
These theoretical 'experiments' must follow the rules of quantum field theory and do not make resort to any specific model. As in true experiments, 
one has to deal with many systematic effects and many limitations of the setup reduce the precision of the final result. Conversely, one of the great advantage 
of this setup is that the freedom of the values of free parameters allows to study various unreal universes, for instance where light quarks are much 
heavier than in Nature or where the light and strange quarks are degenerate.
\subsection{Lattice QCD technology}
One way in which we can realize such 'experiments' is through numerical simulations on a supercomputer. Finite volume of space-time is discretized making the problem
finite dimensional which can be handled by computers. The non-zero lattice spacing regularizes ultraviolet divergences and therefore to any physical observable corresponds
a higly dimensional, finite integral which can be estimated using Monte Carlo integration. Correctly distributed samples, or configurations, are generated by a numerical
implementation of a Markov chain. For given values of the bare parameters of the QCD action it is sufficient to generate the configurations only once, any desired
physical observable can be estimated afterwards. Everybody is invited to design his own theoretical 'experiment' and realize it using his favorite set of gauge configurations.
\subsection{Limitations and solutions}
As usual systematic effects affect the precision of the final result. Below we mention four problems and their possible solutions that were proposed in the last years without 
going into any technical details. In the next section I illustrate how these advancements translate into improvement of the precision of the final result, the second moment
of the pion distribution amplitude.
\begin{itemize}
\item physical pion mass

For very long lattice QCD simulations were limited to the region of parameter space where the quark masses were relatively heavy, much heavier than 
the physical up and down quarks. That was due to the fact that the numerical cost of inversions of the Wilson-Dirac operator needed during 
the generation of gauge field configurations was growing beyond practical limits when the quark masses were lowered. The adoption of sophisticated 
preconditioners, mainly deflative and multigrid methods \cite{preconditioners} which have been developed in the context of problems described by partial 
differential equations, changed that situation. They rely on the idea of separating local and global contributions to the solution and handling them differently. 
In the case of lattice QCD, the part of the solution corresponding to the small eigenvalues of the Wilson-Dirac operator is found in a basis which 
is precalculated during the setup phase of the solver whereas the remaining part corresponding to larger eigenvalues is found iteratively. Moreover the two parts 
can be approximated on lattices with different resolutions. As a result the time to solution became almost independent of the quark mass, allowing simulations
at or even below the physical point \cite{physical_point}.
\item growing autocorrelations

It was found in simulations with periodic boundary conditions for the gauge fields that autocorrelation times of the Markov process grow much faster than expected 
when the lattice spacing is decreased \cite{autocorrelations}. It was later understood that this is related to the freezing of fluctuations of the topological 
charge on consecutive gauge field configurations
generated by the process \cite{autocorrelations2}. In practical terms that meant that simulations with lattice spacing smaller than 0.05 fm were untrustful because of the unestimated autocorrelation
times and therefore because of the inability to correctly estimate the statistical uncertainties of the final results. As a remedy open boundary conditions in the time direction
were proposed \cite{openboundaries} and it was found that indeed the autocorrelation times for such small lattice spacings became manageable, however still quite large \cite{cls}.
\item signal-to-noise ratio problem

Observables involving moving hadrons are known to be much noisier than those with hadrons at rest. For the latter a whole industry of appropriate smearing techniques was
built allowing to construct interpolating operators which create wavefunctions with optimal overlaps with the hadron's ground state. It was only recently that these
techniques were extended to moving hadrons and it was shown that an interpolating operator with the so-called momentum smearing provides a substantially better overlap 
with the wavefunction of a moving particle \cite{momentum_smearing}.
\item structure functions as light-cone correlations

In the case of hadron structure functions calculations performed so far were limited to moments of structure functions, because the structure functions
themselves are defined in terms of light-cone correlations which are not directly accessible on a Euclidean lattice. It was realized only recently by Ji \cite{ji}, 
that a similar information can be extracted from purely space-like correlations if the structure functions are rewritten in the hadron's infinite momentum 
reference frame. This allows to study directly the $x$-dependence of structure functions by Monte Carlo simulations. Since then, several practical 
implementations of Ji's idea have been put forward and first results were published \cite{lamet}.
\end{itemize}
In the next section I define the pion distribution amplitude and its second moment and I demonstrate how each of the above improvements contributes to the precision of the final result.
\section{Pion distribution amplitude}
%
%
Pion distribution amplitude is the quantum amplitude that the pion moving with momentum $P$ is built of a pair of quark and antiquark moving with momentum $xP$ and $(1-x)P$ 
respectively \cite{radyushkin}. It is used, for instance to describe pion photoproduction, where two virtual photons annihilate and produce a pion. Due to factorization 
the transition form factor can be expressed as a product of two parts: photons provide a hard scale and their cross-section can be reliably evaluated with perturbation theory, 
whereas the missing nonperturbative information about the leading pion quark content is contained in the $x$-dependent distribution amplitude. By construction the pion
distribution amplitude is scheme and scale dependent, but is process independent. Experimentally such process was measured by the BaBar \cite{babar} and Belle \cite{belle} experiments.

In mathematical terms the pion distribution amplitude appears on the right hand side of equation Eq.\eqref{eq. definition} and is defined by the following non-local matrix element \cite{braun}
\begin{equation}
\langle 0 | \bar{d}(z_2 n) \slashed n \gamma_5 [z_2 n, z_1 n] u(z_1 n)| \pi(p) \rangle =
i f_{\pi}(p \cdot n) \int_0^1 dx e^{-i(z_1x+z_2(1-x)) p \cdot n} \phi_{\pi}(x,\mu^2)
\label{eq. definition}
\end{equation}
If we neglect isospin breaking effects then $\phi_{\pi}(x)$ becomes symmetric under the interchange 
of momentum fraction 
$\phi_{\pi}(x, \mu^2) = \phi_{\pi}(1-x,\mu^2)$ 
and therefore moments of the momentum fraction difference $\xi = x - ( 1- x)$
carry all the interesting information, 
or equivalently, one can exand in the Gegenbauer polynomials basis and study the corresponding $a^{\pi}_{2}(\mu)$ moment,
\begin{equation}
\langle \xi^n \rangle = \int_0^1 dx (2x - 1)^n \phi_{\pi}(x,\mu^2), \qquad
\phi_{\pi}(x,\mu^2) = 6u(1-u)\Big[ 1+ \sum_n a^{\pi}_{2n}(\mu) C^{3/2}_{2n}(2u-1)\Big],
\end{equation}
The nonlocal operator of Eq.\eqref{eq. definition} can be Taylor 
expanded and expressed in terms of local operators with derivatives
\begin{equation}
\bar{d}(z_2 n) \slashed n \gamma_5 [z_2 n, z_1 n] u(z_1 n) = \sum_{k,l=0}^{\infty}\frac{z_2^k z_1^l}{k!l!} n^{\rho}
n^{\mu_1} \dots n^{\mu_{k+l}} \mathcal{M}^{(k,l)}_{\rho,\mu_1,\dots,\mu_{i+1}}
\end{equation}
where
\begin{equation}
\mathcal{M}^{(k,l)}_{\rho,\mu_1,\dots,\mu_{k+l}} = \bar{d}(0) \overleftarrow{D}_{(\mu_1} \dots \overleftarrow{D}_{\mu_{k}}
\overrightarrow{D}_{\mu_{k+1}} \dots \overrightarrow{D}_{\mu_{k+l}} \gamma_{\rho)}\gamma_5 u(0)
\end{equation}
Consequently, moments $\langle \xi^n \rangle$ can be extracted from expectation values of local operators with derivatives
\begin{equation}
i^{k+l} \langle 0 | \mathcal{M}^{(k,l)}_{\rho,\mu_1,\dots,\mu_{k+l}} | \pi(p) \rangle
 = i f_{\pi} p_{(\rho} p_{\mu_1} \dots p_{\mu_{k+l})} \langle x^l ( 1- x)^k \rangle
\end{equation}
which, apart from some renormalization complications, are easily accessible in Monte Carlo simulations. 
%
%
The strategy to estimate nonperturbatively the second moment of $\phi_{\pi}(x,\mu^2)$ is the following. Due to rotational symmetry breaking 
in lattice QCD we need to study two local operators
\begin{equation}
\mathcal{O}^{\pm}_{\rho \mu \nu}(x) = \bar{d}(x) \Big[ \overleftarrow{D}_{(\mu} \overleftarrow{D}_{\nu}
\pm 2 \overleftarrow{D}_{(\mu} \overrightarrow{D}_{\nu} 
+ \overrightarrow{D}_{(\mu} \overrightarrow{D}_{\nu} \Big] \gamma_{\rho)}\gamma_5 u(x), \label{eq. operators A} 
\end{equation}
The renormalized second moment and its equivalent second Gegenbauer moment are expressed in terms of ratios $R^{\pm}$ and $\zeta_{ij}$, 
\begin{equation}
\langle \xi^2 \rangle^{\overline{\textrm{MS}}} = \zeta_{11} R^- + \zeta_{12} R^+, \qquad
a_2^{\overline{\textrm{MS}}} = \frac{7}{12} \Big[ 5 \zeta_{11} R^- + (5\zeta_{12} - \zeta_{22} ) R^+ \Big].
\end{equation}
$\zeta_{ij}$ are renormalization constants which we estimate nonperturbatively in a separate calculation.
The ratios $R^{\pm}$ are fitted to the ratios of correlation functions $C_{\rho}(t,\mathbf{p})$ and $C^{\pm}_{\rho \mu \nu}(t,\mathbf{p})$ 
which are correlation functions of the axial-vector current $J_{\gamma_5}$ and of operators Eq.\eqref{eq. operators A}
\begin{equation}
R^{\pm}_{\rho \mu \nu, \sigma}(t, \mathbf{p}) = \frac{C^{\pm}_{\rho \mu \nu}(t,\mathbf{p})}{C_{\sigma}(t,\mathbf{p})}.
\end{equation}
namely
\begin{equation}
C_{\rho}(t,\mathbf{p}) = a^3 \sum_{\mathbf{x}} e^{-i \mathbf{p} \mathbf{x}} \langle \mathcal{O}_{\rho}(\mathbf{x},t)
J_{\gamma_5}(0) \rangle \ \textrm{ and } \
C^{\pm}_{\rho \mu \nu}(t,\mathbf{p}) = a^3 \sum_{\mathbf{x}} e^{-i \mathbf{p} \mathbf{x}} \langle \mathcal{O}^{\pm}_{\rho \mu \nu}(\mathbf{x},t)
J_{\gamma_5}(0) \rangle.
\label{eq. correlation functions}
\end{equation}
%
%
%
The RQCD collaboration aims at estimating $\langle \xi^2 \rangle^{\overline{\textrm{MS}}}$ and $a_2^{\overline{\textrm{MS}}}$ 
with a precision of less than 10\%. To this goal we employed all four of the mentioned improvements:
\begin{itemize}
\item we use the set of gauge field configurations generated by the Coordinated Lattice Simulations consortium \cite{cls} which covers ensembles with lattice spacing ranging from 0.04 fm up to 0.086 fm and pion masses from the physical pion mass up to the mass of 420 MeV. The estimates of the second moment include the data point with physical quark masses,
\item CLS ensembles use open boundary conditions in the time direction for the gauge fields in order to keep the autocorrelation times of the Markov chains under control. This enables us
to use gauge field ensembles with lattice spacing of 0.04 fm which improves the precision of the chiral and continuum extrapolations,
\item in order to increase the statistical precision momentum smearing \cite{momentum_smearing} was
used to estimate the correlation functions Eq.\eqref{eq. correlation functions} for non-zero momenta $\mathbf{p}$. An example of a plateau is shown on figure \ref{fig. plateau} \cite{moment}.
\begin{figure}
\begin{center}
\includegraphics[width=0.45\textwidth]{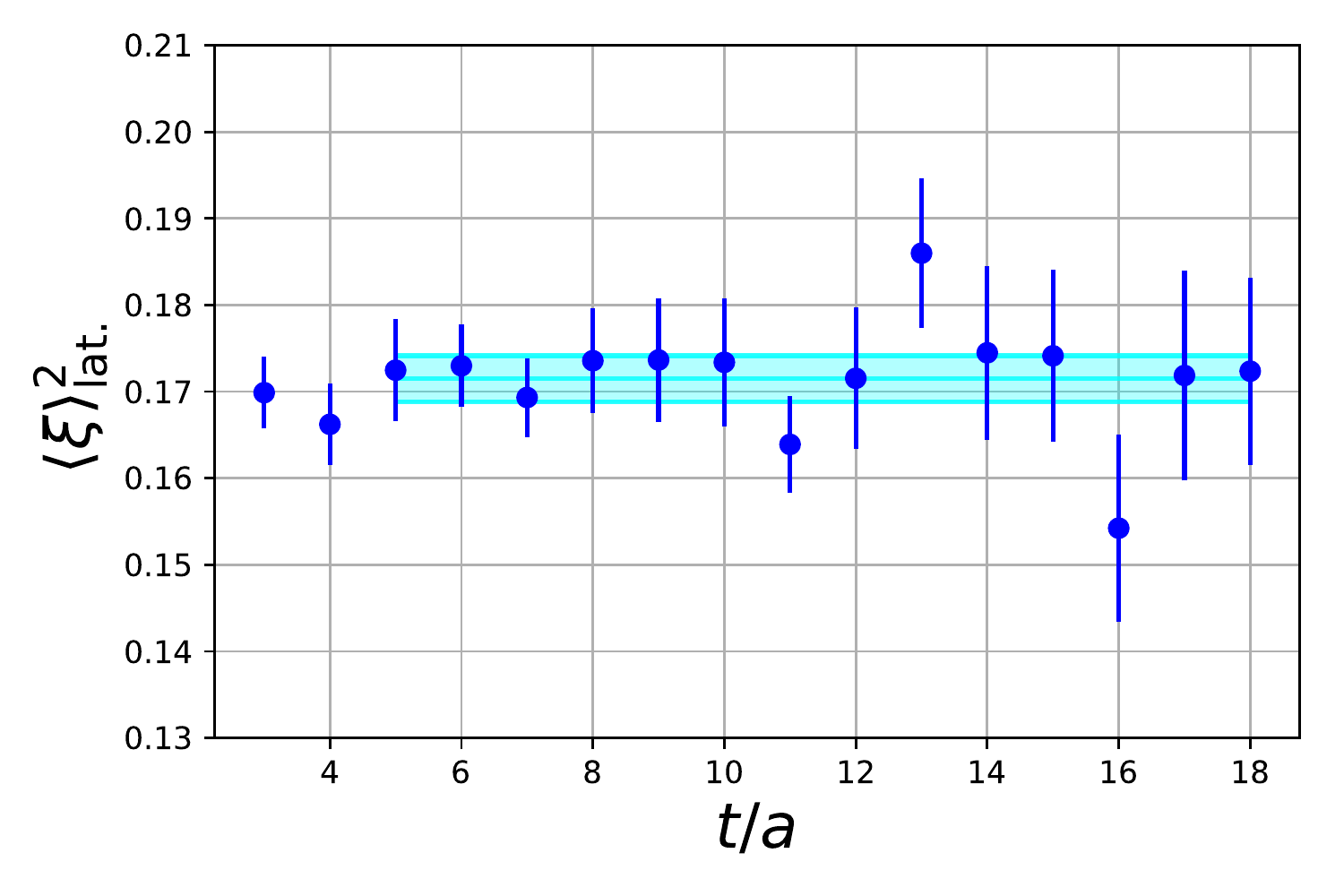}
\includegraphics[width=0.45\textwidth]{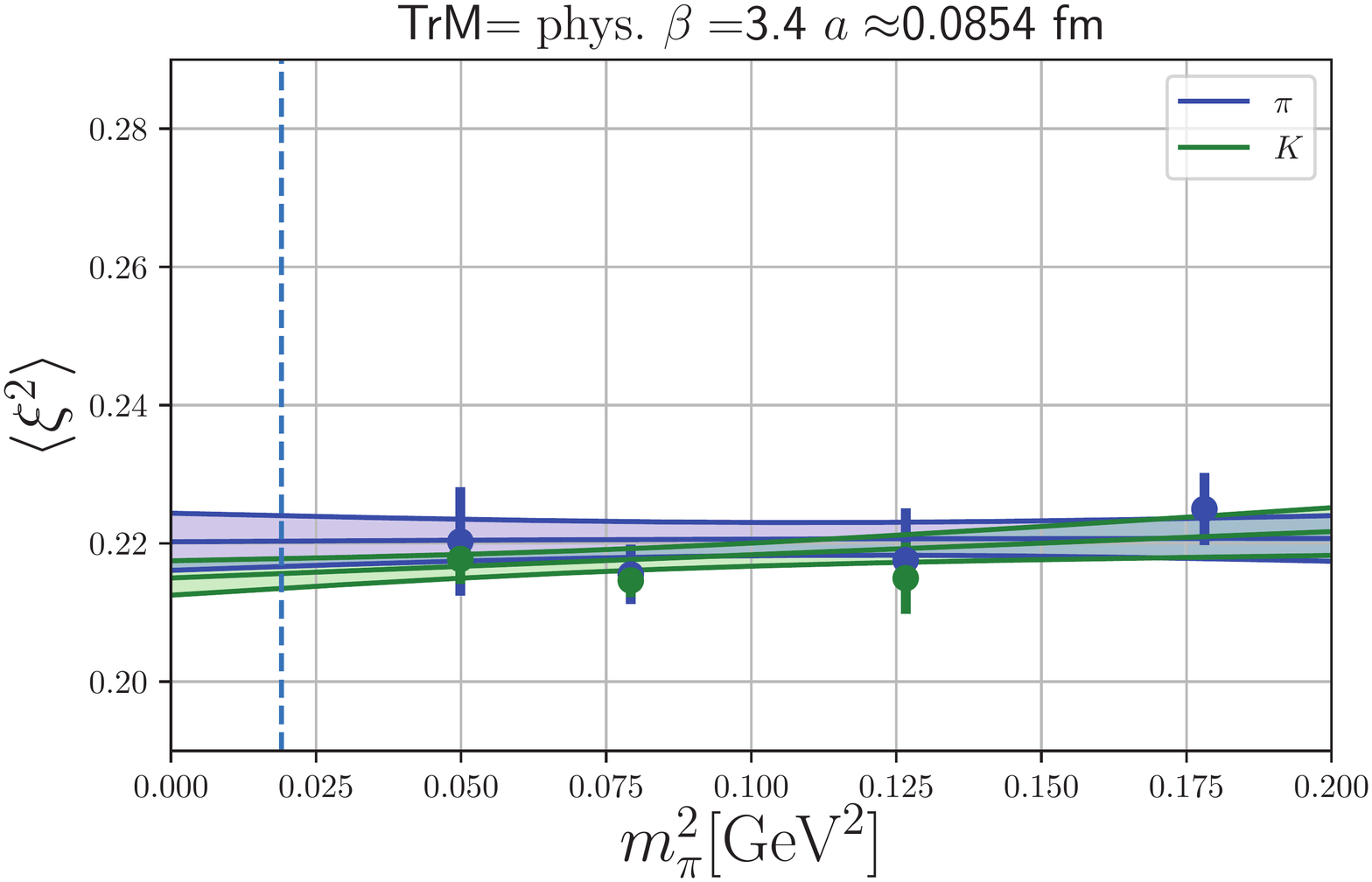}
\caption{Left panel: example of the plateau exhibited by the ratio $R^{+}_{\rho \mu \nu, \sigma}(t, \mathbf{p})$. Momentum smearing allowed to considerably decrease the statistical noise. 
Right panel: example of the extrapolated pion distribution amplitude second moment as a function of the pion mass for $a = 0.086$ fm. \label{fig. plateau}}
\end{center}
\end{figure}
\end{itemize}
Estimates from all available CLS gauge field ensembles are analysed simultaneously in a combined chiral and continuum extrapolations fit. Moreover, since the CLS ensembles
feature two light and one strange dynamical quark, one can estimate pion and kaon distribution second moments together. The dependence on the pion mass 
is determined from continuum chiral perturbation theory and leaves two free parameters $\overline{A}$, $\delta A$, whereas the quadratic discretization effects 
are parametrized by three additional parameters $c_i$. The pion and kaon second moment fit formulae read ($\alpha =$ pion or kaon)
\begin{equation}
\langle \xi^2 \rangle_{\alpha} = \big( 1 + c_0 a + c_1 a \overline{M}^2 + c^{\alpha}_{2} a \delta M^2 \big) 
 \langle \xi^2 \rangle_0 + \overline{A} \ \overline{M}^2 - 2 \delta A \delta M^2
\end{equation}
with
\begin{equation}
\overline{M}^2 = \frac{2m^2_K + m_{\pi}^2}{3}, \qquad \qquad \delta M^2 = m^2_K - m_{\pi}^2
\end{equation}
A single example of the fit is shown on the left panel of figure \ref{fig. plateau} for the lattice spacing of 0.086 fm. Preliminary results for the
continuum extrapolated value of the second moment of the pion distribution amplitude look very promising and the final result will be published soon.

Alternatively, the recent proposal of Ji \cite{ji} suggests that the light-cone correlations which define the hadron structure functions
on the light-cone can be extracted from purely space-like correlations of hadrons moving at large momentum.  The latter are
 calculable in Euclidean formulation of lattice QCD, and the in the infinite momentum limit one recovers the light-cone distributions.
The framework of Large Momentum Effective Theory allows to systematically calculate corrections coming from large but finite hadron 
momentum achievable on the lattice. Current applications to the pion distribution amplitude show a quantitative agreement with other methods \cite{lamet_pion}.
\section{Conclusions}
Lattice QCD provides non-perturbative, \emph{ab initio} results for many observables and hadron structure functions in particular. 
Continous progress in the computational strategies and algorithms allows to improve our estimates for phenomenologically relevant 
observables. On one hand, statistical and systematic uncertainties of already studied quantities can be made smaller and smaller. 
We mentionned advancements in solver techniques allowing simulations at the physical quark masses and new boundary conditions which 
help to control Markov chain autocorrelations. On the other hand, new computational tools such as the Large Momentum Effective Theory 
give access to observables, which were not studied so far on the lattice. I illustrated that using the pion distribution amplitude, 
for which the systematic effects are becoming under control as far as the second moment is concerned, whereas a new method to 
extract the full $x$-dependence of structure functions is providing already good qualitative results.

\begin{acknowledgement}
The Author whishes to thank all colleagues from the RQCD and CLS collaborations. This work has been supported by the Deutsche 
Forschungsgemeinschaft (SFB/TRR-55) and the Polish National Science Center (NCN grant number UMO-2016/21/B/ST2/01492).
Computer time allocations at the following centers are gratefully acknowledged: the Interdisciplinary Centre for Mathematical 
and Computational Modelling (ICM) of the University of Warsaw (grant No. GA67-12, GA69-20, GA71-26), HLRS (Universität Stuttgart), 
JSC (Forschungszentrum Jülich), and LRZ (Bayerische Akademie der Wissenschaften).
\end{acknowledgement}
%

\begin{thebibliography}{00}
%
%
\bibitem{resonances}
R.~Briceno, "Resonances from lattice QCD", EPJ Web of Conferences \textbf{175}, 01016 (2018) 
\bibitem{scattering}
M.~Hansen, review talk, Lattice2015 conference, arXiv:1511.04737
\bibitem{nuclear}
M.~Savage, review talk, Lattice2016 conference, Z.~Davoudi, review talk, Lattice2017 conference, 10.1051/epjconf/201817501022
\bibitem{preconditioners}
A.~Frommer, review talk, Lattice2014 conference
\bibitem{physical_point}
S.~Durr~\textit{et al.}, 10.1007/JHEP08(2011)148, S.~Durr~\textit{et al.}, Phys.Lett. B \textbf{701} 265 (2011), A. Abdel-Rehim
\textit{et al.} (ETM Collaboration), Phys. Rev. D \textbf{95}, 094515 (2017)
\bibitem{autocorrelations}
S.~Schaefer, Nucl.Phys. B \textbf{845}, 93 (2011)
\bibitem{autocorrelations2}
M.~Luscher~\textit{et al.}, 10.1007/JHEP07(2011)036
\bibitem{openboundaries}
M.~Luscher~\textit{et al.}, 10.1016/j.cpc.2012.10.003
\bibitem{cls}
M.~Bruno~\textit{et al.}, 10.1007/JHEP02(2015)043
\bibitem{momentum_smearing}
G.~Bali~\textit{et al.}, Phys. Rev. D \textbf{93}, 094515 (2016)
\bibitem{ji}
X.~Ji, Phys. Rev. Lett. \textbf{110}, 262002 (2013)
\bibitem{lamet}
X.~Ji~\textit{et al.}, Phys. Rev. Lett. \textbf{111}, 112002 (2013), review talk by C. Monahan "Recent Developments in x-dependent Structure Calculations", Lattice2018 conference
\bibitem{radyushkin}
A.~Efremov~\textit{et al.}, Phys. Lett. B. \textbf{94}, 245 (1980)
\bibitem{babar}
B.~Aubert~\textit{et al.}, Phys. Rev. D \textbf{80}, 052002 (2009)
\bibitem{belle}
S.~Uehara~\textit{et al.}, Phys. Rev. D \textbf{86}, 092007 (2012)
\bibitem{braun}
V.~Braun~\textit{et al}, Phys.Rev. D \textbf{92}, 014504 (2015)
\bibitem{moment}
G.~Bali~\textit{et al.}, Phys.Lett. B \textbf{774}, 91 (2017)
\bibitem{lamet_pion}
J-H.~Zhang~\textit{et al.}, Phys.Rev. D \textbf{95} 094514 (2017), V.~Braun~\textit{et al}, Eur.Phys.J. C \textbf{78}, 217 (2018), G.~Bali~\textit{et al.},  arXiv:1807.06671
\end{thebibliography}
%
%

\end{document}